# ADVANCED TRANSPORT OPTIONS FOR THE DYNAMIC ADAPTIVE STREAMING OVER HTTP

*Christian Timmerer[†,‡] and Alan Bertoni[†]*

[†]Alpen-Adria-Universität Klagenfurt, Institute of Information Technology (ITEC), Austria
{*firsname.lastname*}@itec.aau.at

[‡]bitmovin Inc., Palo Alto, USA and Klagenfurt, Austria, https://bitmovin.com/
christian.timmerer@bitmovin.com

**ABSTRACT**

*Multimedia streaming over HTTP is no longer a niche research topic as it has entered our daily live. The common assumption is that it is deployed on top of the existing infrastructure utilizing application (HTTP) and transport (TCP) layer protocols as is. Interestingly, standards like MPEG's Dynamic Adaptive Streaming over HTTP (DASH) do not mandate the usage of any specific transport protocol allowing for sufficient deployment flexibility which is further supported by emerging developments within both protocol layers. This paper investigates and evaluates the usage of advanced transport options for the dynamic adaptive streaming over HTTP. We utilize a common test setup to evaluate HTTP/2.0 and Google's Quick UDP Internet Connections (QUIC) protocol in the context of DASH-based services.*

## 1. INTRODUCTION

Adaptive multimedia streaming over-the-top of the existing infrastructure using HTTP is a major driver for innovation within both industry and academia. The MPEG standard Dynamic Adaptive Streaming over HTTP (DASH) provides interoperable representation formats in terms of media presentation description (MPD) and segments based on the ISO base media file format and MPEG-2 transport stream [1]. Interestingly, the standard mandates the usage of HTTP-URLs for locating segments but not how they are actually delivered to the client. The general assumption is that a standard HTTP infrastructure is used which is deployed on top of TCP for the delivery of both MPD and segments. In practice, however, various (transport) protocols could be used such as in 3GPP which specifies DASH over (e)MBMS/FLUTE in a mobile broadcast environment [2]. Another option for DASH is the recently proposed version 2 of HTTP – written as HTTP/2.0 – which is based on Google's SPDY [3]. HTTP/2.0 comes with an interesting pool of features that could be exploited in the context of DASH. For instance, Wei and Swaminathan propose k-push (i.e., k segments are pushed to the client using one request) to reduce both latency and the number of segment requests using the HTTP/2.0 server push feature [4]. Others use HTTP chunked transfer encoding to achieve similar latency requirements [5][6].

While HTTP/2.0 is tightly coupled with TCP, earlier versions of HTTP actually do not mandate the usage of TCP although almost all implementations assume TCP to be used, specifically its means for reliable transport. The performance of TCP for media streaming applications has been analytically assessed in [7] concluding that the bandwidth requirement is about twice the media bitrate. Various improvements – both at the HTTP and the TCP layer – have shown significant performance gain, specifically when adopting persistent connections and pipelined requests as defined within HTTP/1.1 [8]. These features definitely provide a performance boost but suffers from the Head-of-Line (HoL) blocking problem and together with TCP's streaming inflexibility this has lead to ad-hoc developments such as Google's Quick UDP Internet Connections (QUIC) protocol [9] and also SPDY; the latter being turned into HTTP/2.0 at the time of writing this paper. While the performance of HTTP/2.0 in the context of DASH has been assessed already (i.e., compared with selected features of HTTP/1.0 and HTTP/1.1) its combination with QUIC has not yet been evaluated to the best of our knowledge.

The aim of this paper is to provide a *baseline performance assessment of DASH-based services with advanced transport options both at the application and transport layer*. At the application layer, we investigate the usage of SPDY/HTTP/2.0 and HTTP/1.1 (with persistent connection and pipelined requests enabled); at the transport layer, we consider TCP and QUIC. For the actual evaluation we examine the protocol overhead, link utilization, and adaptation performance for the following combinations: *(a)* HTTP/2.0 over TCP, *(b)* HTTP/2.0 over SSL (and TCP), *(c)* HTTP/1.1 over QUIC, and *(d)* SPDY over QUIC. Please note that HTTP/2.0 and SPDY share the same principles despite minor format differences but this shall not impact its performance. In this paper we adopt MPEG-DASH for the actual streaming format but results are also applicable for other formats sharing the same principles (e.g., Apple HTTP Live Streaming). The evaluation setup is compliant with [3] to enable cross-validation with both results, e.g., when targeting future enhancements.

The remainder of the paper is as follows. Section 2 provides a brief overview of MPEG-DASH and Section 3 describes advanced transport options, i.e., HTTP/2.0 and QUIC. Section 4 describes the experimental evaluation setup and discusses the evaluation results. The paper concludes with Section 5 highlighting also future work.

## 2. MPEG-DASH

MPEG-DASH and related formats sharing the same principles (e.g., Adobe HDS, Apple HLS, Microsoft Smooth Streaming) enable adaptive HTTP streaming by providing multiple, time-aligned versions (e.g., different bitrate, resolution, codec, language) of segmented media files (e.g., 2-10 seconds) on ordinary Web servers which clients individually request in a dynamic and adaptive way depending on its usage environment (e.g., available bandwidth, display resolution, codec support, language preference of the user). Sodagar gives a most recent overview of the MPEG-DASH standard [1] which provides a specification for the MPD and segment formats based on ISO base media file format and MPEG-2 transport stream.

A major requirement of the standard was to support the usage of standard Web servers without the need for any media- or streaming-specific extensions to enable reuse of the existing infrastructure deployed for the provisioning and delivery of regular Web traffic. The common assumption is that the intelligence is solely within the client implementation which requests segments – as described in the MPD – based on its context conditions. This adaptive client behavior and the supported media codecs are not normatively defined within the standard. The DASH Industry Forum (DASH-IF; http://dashif.org/) provides interoperability points going beyond the MPEG specification including recommendations for selected media codecs, test vectors, and conformance software.

A detailed state of the art and open issues can be found in the tutorial of Timmerer and Begen [10].

## 3. ADVANCED TRANSPORT OPTIONS: SPDY/HTTP/2.0 AND QUIC

### 3.1. SPDY and HTTP/2.0

This section describes HTTP/2.0 which is based on Google's SPDY protocol and at the time of writing of this paper available as Internet draft by the IETF [12].

The protocol is mandating the Transmission Control Protocol (TCP) and maintains a single persistent connection for each session. During a session, multiple streams can be opened between the client and the server in full-duplex mode. Typically, only one HTTP/2.0 connection between a server and a client exists until the client navigates to another server. The servers should leave connections open as long as possible until a given threshold timeout or when a client initiates a connection close.

The advantage of HTTP/2.0 is that it is fully compatible with HTTP/1.1 and can be integrated as a session layer between HTTP and TCP, hence, enabling incremental deployment. The HTTP request will be mapped into a HTTP/2.0 frame and vice versa for the HTTP response. Additionally, it is also possible to send multiple requests in parallel to support pipelining. Therefore, HTTP/2.0 offers an interface for HTTP, which simplifies its integration for already existing HTTP applications. After this handover from HTTP/1.1 to HTTP/2.0 the whole communication will be handled on the HTTP/2.0 framing layer until a response arrives which will be passed to the HTTP/1.1 layer.

Google further developed SPDY which is currently available as Draft 3.1 and still maintains two frame types for control and data frames but with similar functionality as within HTTP/2.0.

### 3.2 QUIC

Quick UDP Internet Connections (QUIC) is an experimental, UDP-based transport layer network protocol, which aims at reduced connection latency, congestion control, multiplexed/pipelined requests without head-of-line blocking, FEC, and connection migration [13].

During the connection establishment, the client speculatively assumes to have acceptable cryptographic credentials for at least a preliminary encryption of a request. In case the server does not accept the credentials, additional negotiations may be needed but, conceptually, all handshakes have a zero-RTT in QUIC.

The variable length (2-19 bytes) packet header comprises public/private flags, connection identifier, version information, sequence number, and FEC data. Streams are independent sequences of bi-directional data packets, which can be created both by the client and the server. For the congestion control, QUIC comes with two different approaches: (a) to mimic the TCP CUBIC algorithm, and (b) an inter-arrival scheme based on WebRTC.

SPDY and QUIC are designed to work independent from each other but when SPDY is implemented over QUIC, the QUIC layer handles most of the stream management. In particular, SPDY streams IDs are replaced by QUIC stream IDs without explicit framing and the data sent over the QUIC stream simply consists of SPDY headers followed by the body.

## 4. EXPERIMENTAL EVALUATION

### 4.1 Evaluation Setup

The test content for the evaluation is Big Buck Bunny which has been encoded and segmented into 14 representations using x264 and MP4Box, respectively. The segment size is 2 seconds and the bitrate varies from 100 to 4,500 kbps as follows: 100, 200, 350, 500, 700, 900, 1100, 1300, 1600, 1900, 2300, 2800, 3400, and 4500. We use a constant frame rate of 30 fps and constant resolution of 640x360 pixels as we are mainly focusing on changes in the actual bitrate. Therefore, the resolution is not that important.

**Table 2. Overhead Analysis for HTTP/2.0 and QUIC.**

| Protocol Stack | Overhead | |
|---|---|---|
| | **HTTP/2.0** | **QUIC** |
| **Transport Layer** | TCP (32 bytes) | UDP (8 bytes) |
| **Network Layer** | IP (20 bytes) | IP (20 bytes) |
| **Data Link Layer** | Ethernet (14 b) | Ethernet (14 b) |
| **MTU [bytes]** | 1,514 | 1,2,42 |
| **Total Overhead [%]** | 4.36% | 3.38% |

**Table 1. Average Link Utilization at Different Round-Trip-Time.**

| Protocol | Link Utilization [%] | | |
|---|---|---|---|
| | **0ms** | **50ms** | **150ms** |
| **HTTP/2.0 over TCP** | 95.3 | 92.9 | 88.4 |
| **HTTP/2.0 over SSL** | 95.1 | 92.6 | 88.0 |
| **HTTP/1.1 over QUIC** | 94.0 | 91.8 | 87.2 |
| **SPDY over QUIC** | 93.9 | 91.7 | 87.2 |

The test environment comprises a sever component (hosting HTTP and QUIC servers) and the DASH client connected through a network emulator responsible for bandwidth shaping (token bucket filter and traffic control program) and network delay emulation (netem program) [3]. The server hosts a standard Apache Web server (v2.4.7) with nghttp2 proxy for the HTTP/2.0 delivery. Moreover, it runs the QUIC prototype server, which supports versions 15 to 19 of the QUIC protocol, to deliver media content using HTTP/1.1 directly over QUIC or using SPDY over QUIC. The DASH client is based on the QTSamplePlayer that comes with libdash which has been enhanced with SPDY/HTTP/2.0/QUIC capabilities and a simple adaptation logic referred to DASH-JS [11].

For each subsequent evaluation, five runs have been conducted and the mean value is presented. Note that differences between individual runs are so marginal that we refrain from showing confidence intervals.

### 4.2 Protocol Overhead

The protocol overhead is first computed based on the underlying specifications and summarized in Table 2. HTTP is based on TCP, which introduces an overhead of 20 bytes for the TCP header and additional 12 bytes for the optional header fields. QUIC is based on UDP, which introduces an overhead of 8 bytes. The remaining overhead is the same for both approaches, e.g., using 20 bytes for the IP header and additional 14 bytes for the Ethernet at the link layer.

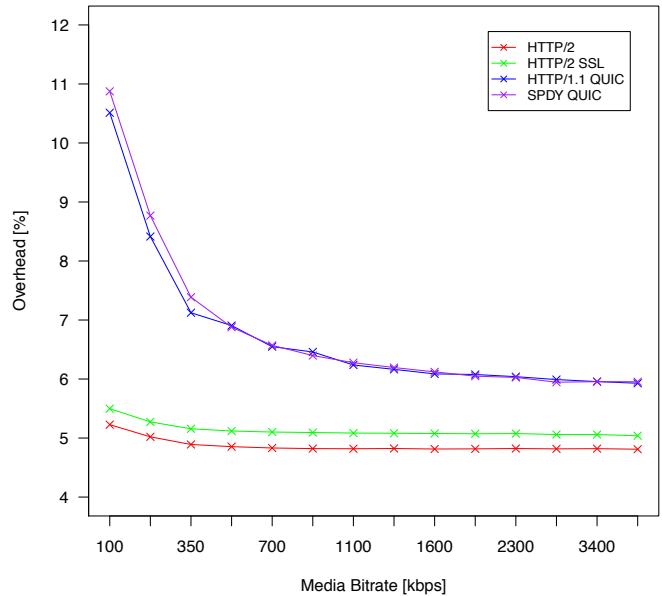

**Figure 1. Protocol Overhead at Different Media Representations (Bitrate).**

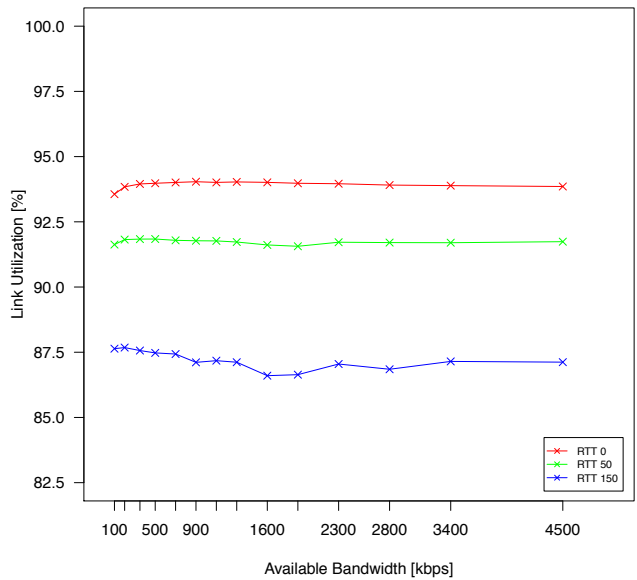

**Figure 2. Link Utilization of SPDY over QUIC.**

HTTP/2.0 and QUIC adopt an additional framing layer above TCP and UPD. For HTTP/2.0, each frame has an 8-byte header to carry the length, stream identifier, type and corresponding flags. For QUIC, the frame header does not have a fixed length but varies between 2 and 19 bytes.

The actual protocol overhead is finally measured for the 14 different representations in a DASH scenario as $1 - \frac{media\ bytes}{total\ bytes\ received}$ and depicted in Figure 1. The horizontal axis shows the quality level of the encoded representation

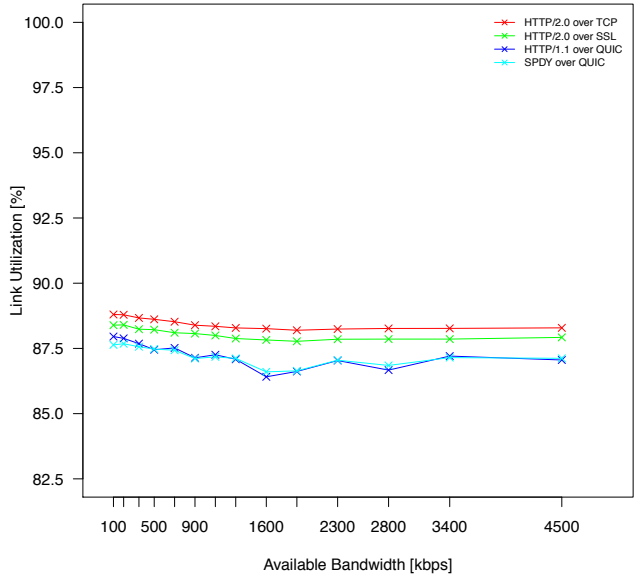

**Figure 3. Comparison of Link Utilization with RTT 150ms.**

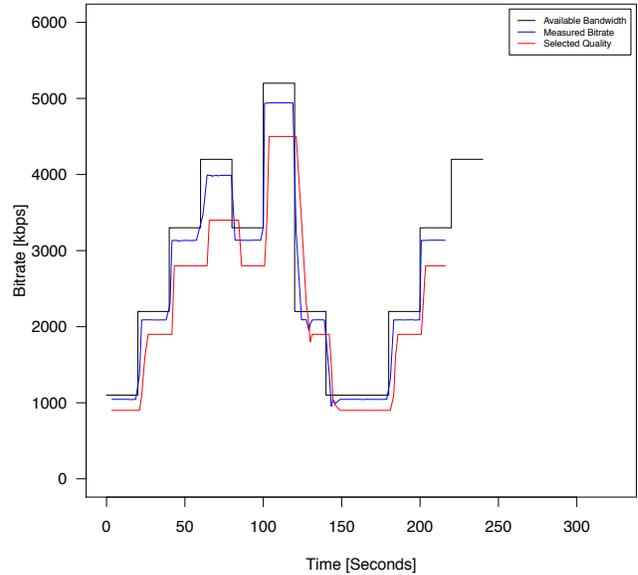

**Figure 4. Adaptation Performance of SPDY over QUIC with RTT 0ms.**

while the vertical axis shows the protocol overhead in percentage.

In general, the overhead is below 10% except for QUIC and very low bitrates at 100 kbps. However, QUIC always comes with a higher overhead than HTTP/2.0 over TCP. This result is counter-intuitive since QUIC is running over UDP, which has a slightly lower protocol overhead than TCP. Furthermore, keep in mind that QUIC provides a multiplexed stream protocol on top of UDP and security comparable with SSL. Comparing the solutions providing encryption, the average overhead of QUIC is about 1.65% higher than HTTP/2.0 over SSL.

### 4.3 Link Utilization

The link utilization has been tested with all representations using different round trip times (RTTs) of 0ms (local area networks), 50ms (fixed line/wired network), and 150ms (wireless/mobile network). The actual link utilization is calculated as a ratio of the effective throughput and the available bandwidth. For each individual run, the bandwidth is restricted to the bitrate of the corresponding representation. As expected, the higher the RTT, the lower the link utilization but in all cases it is >80% as shown in Table 1.

Figure 2 shows the link utilization of SPDY over QUIC for the different RTTs and the given available bandwidth. The results are stable over the bandwidth and similar for the other protocol combinations. A comparison of the link utilization with RTT 150ms is depicted in Figure 3. The comparisons of the other RTTs look similar but with a higher link utilization according to the values shown in Table 1. Interestingly, the link utilization is not as stable when using QUIC compared to TCP/SSL configurations.

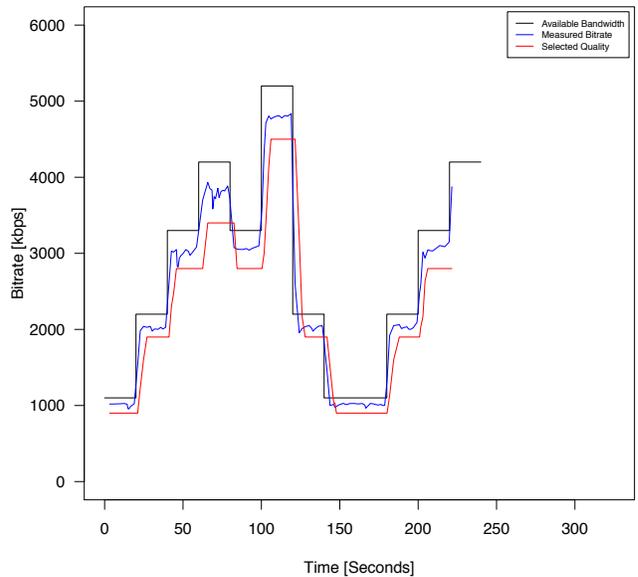

**Figure 5. Adaptation Performance of SPDY over QUIC with RTT 50ms.**

However, QUIC is becoming more stable with decreased RTT (not shown here).

### 4.4 Adaptation Performance

The adaptation performance is evaluated for a given bandwidth trajectory limiting the available bandwidth between 1-5 Mbps for the different protocol combinations. Additionally, the same RTTs as for the link utilization have been used. The results reveal that the adaptation performance – average media throughput – is very similar for the different protocol combinations (>2 Mbps in all cases) and, thus, we focus on SPDY over QUIC for different RTTs.

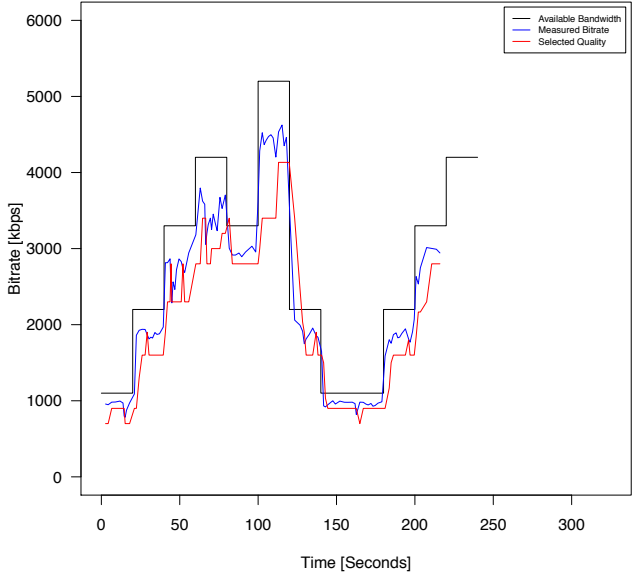

**Figure 6. Adaptation Performance of SPDY over QUIC with RTT 150ms**

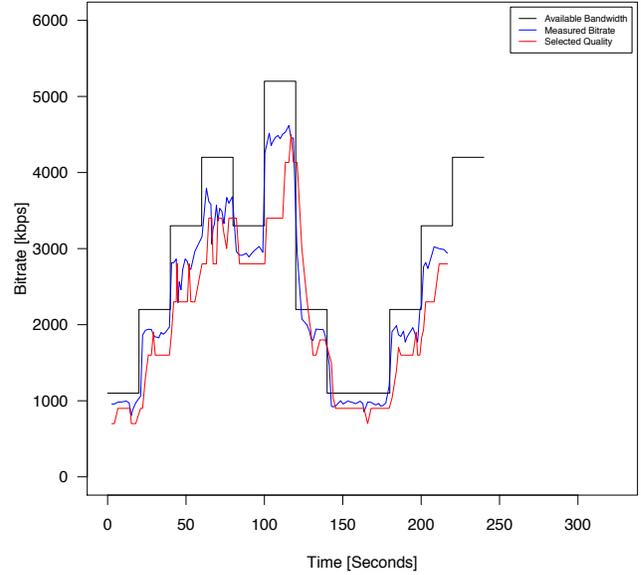

**Figure 7. Adaptation Performance of HTTP/1.1 over QUIC with RTT 150ms.**

For the actual adaptation logic we adopt DASH-JS from [11] which is based on a simple bandwidth estimation as shown in Equation **(1)**.

$$b_n = \frac{w_1 b_{n-1} + w_2 b_m}{w_1 + w_2} \quad (1)$$

where $b_{n-1}$ is the throughput calculated at the $n-1^{th}$ segment, $b_m$ denotes the throughput measured during the download of the $n-1^{th}$ segment, while $w_1$ and $w_2$ are weighting factors that adjust the influence of the recently measured segment download (i.e., $w_1$=0.7 and $w_2$=1.3 according to [11]).

Figure 4 shows the adaptation performance of SPDY over QUIC with RTT 0ms. The black line shows the available bandwidth using the bandwidth shaping within the network emulator. The blue line represents the available bandwidth measured while downloading the actual segments and providing the input for the adaptation logic (i.e., DASH-JS). The red line depicts the output of adaptation logic and corresponds to the selected quality according to the available representations within the MPD.

Figure 5 shows the adaptation performance of SPDY over QUIC with RTT 50ms and Figure 6 with RTT 150. The results reveal that DASH-JS is robust against different RTTs and provides an instant reaction to the available/measured bandwidth. Figure 7 provides the results of the adaptation behavior of HTTP/1.1 over QUIC with RTT 150ms, which is indeed very similar to the results of SPDY over QUIC as shown in Figure 6 and, thus, we can conclude that the adaptation logic does not impact the underlying protocols.

Finally, a comparison of the average media throughput of all protocol combinations for the different RTTs is shown in Figure 8. The black line represents the maximum throughput and comprises the average value of the given bandwidth trajectory (i.e., 2.7 Mbps). The results clearly indicate that all protocol combinations provide roughly the same adaptation performance whereby the media throughput decreases with increasing RTT but always is above 2 Mbps.

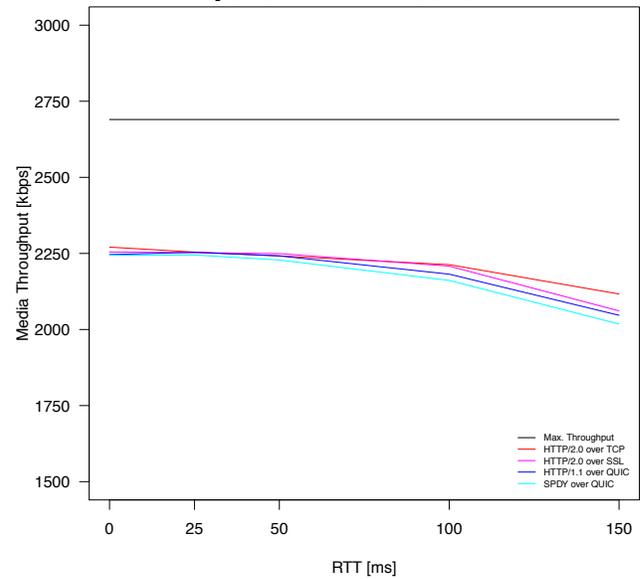

**Figure 8. Comparison of the Adaptation Performance for different RTTs.**

### 4.5. Discussion

In this section we want to discuss the results achieved in this evaluation and compare it with the results reported in a similar study by Mueller et al. focusing on HTTP/2.0 and

SPDY only [3]. In fact, the evaluation setup is identical with that in [3] and, thus, allows for a direct comparison of the results. In principle, we confirm the results of Mueller et al. but have not further investigated HTTP/1.0 as we use consistently HTTP/1.1 including its features like persistent connections and request pipelining. The bandwidth trajectory is different but our results show the same behavior as reported in [3].

In our setup we add QUIC as an alternative to TCP for the actual transport layer protocol, which – together with HTTP/2.0 – eliminates the Head-of-Line blocking problem of pipelined requests in HTTP/1.1. However, using QUIC instead of TCP does not contribute to the overall streaming performance in terms of increased or decreased media throughput at the client.

Interestingly, QUIC, which is based on UDP, comes with a slightly higher overhead than TCP, specifically for low bitrates but is still <10% in all cases and <7% in the majority of the cases.

## 5. CONCLUSIONS AND FUTURE WORK

In this paper we evaluated advanced transport options for the dynamic adaptive streaming over HTTP. Therefore, we evaluated HTTP/1.1/2.0/SPDY over TCP/QUIC using a predefined evaluation setup. In this context, QUIC comes with a slightly higher protocol overhead than TCP but is below 10% except for very low bitrates (≤100kbps). The link utilization decreases with increasing RTT but is always >87% of the available bandwidth and remains stable for different bandwidths. The adaptation algorithm does not have an impact on the transport scheme used but the media throughput decreases with increasing RTT. Thus, results reported in this paper confirm previous results in [3] but provide additional findings for QUIC.

Future work includes studying further advanced transport options such as Akamai's hybrid HTTP/UDP approach – as known as Astraeus – which has been specifically designed for large packet sizes [14]. Therefore, we will investigate different DASH segment sizes and how to combine them with such advanced transport options to increase the overall delivery performance, possibly including Quality of Experience (QoE) aspects.

## 12. REFERENCES

[1] I. Sodagar, "The MPEG-DASH Standard for Multimedia Streaming Over the Internet," *IEEE MultiMedia*, vol.18, no.4, pp.62-67, Apr. 2011. doi: 10.1109/MMUL.2011.71

[2] T. Lomar, M. Slessingar, V. Kenehan, S. Puustien, "Delivering content with LTE Broadcast," *Ericsson Review*, Feb. 2013. http://bit.ly/14ywYy9 (last access: Nov. 2014)

[3] C. Mueller, S. Lederer, C. Timmerer, H. Hellwagner, "Dynamic Adaptive Streaming over HTTP/2.0," 2013 *In Proceedings of IEEE International Conference on Multimedia and Expo (ICME'13)*, San Jose, CA, USA, Jul. 2013. doi: 10.1109/ICME.2013.6607498

[4] S. Wei and V. Swaminathan, "Low Latency Live Video Streaming over HTTP 2.0," *In Proceedings of Network and Operating System Support on Digital Audio and Video Workshop (NOSSDAV'14)*, Singapore, Mar. 2014. doi=10.1145/2578260.2578277

[5] V. Swaminathan, S. Wei, "Low latency live video streaming using HTTP chunked encoding," *In Proceedings of IEEE 13th International Workshop on Multimedia Signal Processing (MMSP'11)*, Hangzhou, China, Oct. 2011. doi: 10.1109/MMSP.2011.6093825

[6] N. Bouzakaria, C. Concolato, J. Le Feuvre, "Overhead and performance of low latency live streaming using MPEG-DASH," *In Proceedings of 5th International Conference on Information, Intelligence, Systems and Applications (IISA'14)*, Chania, Greece, Jul. 2014. doi: 10.1109/IISA.2014.6878732

[7] B. Wang, J. Kurose, P. Shenoy, D. Towsley, "Multimedia Streaming via TCP: An Analytic Performance Study," *ACM Transactions on Multimedia Computing, Communications, and Applications (TOMM)*, vol. 4, no. 2, May 2008. doi=10.1145/1352012.1352020

[8] C. Müller, S. Lederer, C. Timmerer, "An Evaluation of Dynamic Adaptive Streaming over HTTP in Vehicular Environments," *In Proceedings of the 4th Workshop on Mobile Video (MoVid'12)*, Chapel Hill, NC, USA, Feb 2012. doi=10.1145/2151677.2151686

[9] B. Trammell, J. Hildebrand, "Evolving Transport in the Internet," *IEEE Internet Computing*, vol.18, no.5, Sep.-Oct. 2014. doi: 10.1109/MIC.2014.91

[10] C. Timmerer and A. C. Begen, "Over-the-Top Content Delivery: State of the Art and Challenges Ahead," *In Proceedings of the ACM International Conference on Multimedia (MM'14)*, Orlando, FL, USA, Nov. 2014. doi=10.1145/2647868.2654849

[11] B. Rainer, S. Lederer, C. Mueller, C. Timmerer, "A seamless Web integration of adaptive HTTP streaming," *In Proceedings of the 20th European Signal Processing Conference (EUSIPCO'12)*, Bucharest, Romania, Aug. 2012.

[12] M. Belshe, et al., "Hypertext Transfer Protocol version 2.0", draft-ietf-httpbis-http2-14, Oct. 2014, http://tools.ietf.org/search/draft-ietf-httpbis-http2-15.

[13] QUIC, a multiplexed stream transport over UDP, http://www.chromium.org/quic

[14] M. Ponec, A. Alness, "Hybrid HTTP and UDP content delivery," US Patent US20140059168, Feb. 2014. http://www.google.com/patents/US20140059168